\documentclass[manuscript]{aastex}

\shorttitle{M dwarf catalog of LAMOST pilot survey }
\shortauthors{Yi \& Luo et al.}

\begin{document}
\title{M dwarf catalog of LAMOST pilot survey }

\author{Z.P. Yi \altaffilmark{1,2,3,4}, A.L. Luo \altaffilmark{1,3,8}, Y.H. Song \altaffilmark{1,3}, J.K. Zhao\altaffilmark{1,3}, Z.X. Shi \altaffilmark{1,4}, P. Wei\altaffilmark{1,4}, J.J. Ren \altaffilmark{1,4}, F.F. Wang\altaffilmark{1,3},  X. Kong \altaffilmark{1,3},Y.B. Li \altaffilmark{1,3}, P. Du,\altaffilmark{1,2,3},  W. Hou \altaffilmark{1,4}, Y.X. Guo \altaffilmark{1,3}, S. Zhang \altaffilmark{1,4}, Y.H. Zhao\altaffilmark{1,3}, S.W. Sun \altaffilmark{1,3},}
\affil{National Astronomical Observatories, Chinese Academy of Sciences,
    Beijing, 100012, China}
\email{lal@nao.cas.cn, lal@lamost.org}

\author{ J.C. Pan\altaffilmark{2}}
\affil{Shandong University at Weihai, 180 Wenhuaxi Road, Weihai, 264209, China}

\author{ L.Y. Zhang\altaffilmark{5}}
\affil{College of Science\/Department of Physics, Guizhou University, Guiyang, 550025, China;}

\author{A. A. West\altaffilmark{6}}
\affil{Department of Astronomy, Boston University, 725 Commonwealth Ave, Boston,MA 02215, USA}

\and
\author{H.B. Yuan\altaffilmark{7}}
\affil{Kavli Institute for Astronomy and Astrophysics, Peking University, Beijing, 100871 China}

\altaffiltext{1}{National Astronomical Observatories(NAOC), Chinese Academy of Sciences, Beijing 100012, China.}
\altaffiltext{2}{Shandong University at Weihai, 180 Wenhuaxi Road, Weihai, 264209, China.}
\altaffiltext{3}{Key Laboratory of Optical Astronomy, NAOC, Chinese Academy of Sciences, Beijing 100012, China.}
\altaffiltext{4}{University of Chinese Academy of Sciences, 19A Yuquan Road, Shijingshan District, Beijing, 100049, China}
\altaffiltext{5}{College of Science/Department of Physics, Guizhou University, Guiyang 550025, China}
\altaffiltext{6}{Department of Astronomy, Boston University, 725 Commonwealth Ave, Boston, MA 02215, USA}
\altaffiltext{7}{Kavli Institute for Astronomy and Astrophysics, Peking University, Beijing, 100871 China}
\altaffiltext{8}{Correspondence Author}

\begin{abstract}
We present a spectroscopic catalog of 67082 M dwarfs from the LAMOST pilot survey. For each spectrum of the catalog, spectral subtype, radial velocity, equivalent width of H${\alpha}$, a number of prominent molecular band indices and the metal sensitive parameter $\zeta$ are provided . Spectral subtype have been estimated by a remedied Hammer program (Original Hammer: Covey et al. 2007), in which indices are reselected to obtain more accurate auto-classified spectral subtypes. All spectra in this catalog have been visually inspected to confirm the spectral subtypes. Radial velocities have been well measured by our developed program which uses cross-correlation method and estimates uncertainty of radial velocity as well. We also examine the magnetic activity properties of M dwarfs traced by H${\alpha}$ emission line. The molecular band indices included in this catalog are temperature or metallicity sensitive and can be used for future analysis of the physical properties of M dwarfs. The catalog is available on the website \url{http://sciwiki.lamost.org/MCatalogPilot/}.
\end{abstract}


\section{Introduction}
M dwarfs are the most common stars in the Galaxy (Bochanski et al. 2010). Their main sequence lifetime is even longer than the present age of the universe (Laughlin et al. 1997). Therefore, M dwarfs can be used to trace the structure and the evolution of the Milky Way. Besides investigating the properties of the Galaxy, M dwarfs are important for identifying potentially habitable extrasolar planets (Charbonneau et al. 2009). Many previous studies have been completed using catalog of M dwarfs, for instance, tracing the Galactic disk kinematics (Hawley et al. 1996; Gizis et al. 2002; L\'{e}pine et al. 2003; Bochanski et al. 2005, 2007a, 2010), studying the structure of the Galaxy (Reid et al.1997; Kerber et al. 2001; Woolf et al. 2012) and computing the stellar initial mass function (Covey et al. 2008b; Bochanski et al. 2010). To research these scientific topics, some fundamental and preliminary analysis need to be performed in advance, including spectral type classification (Kirkpatrick et al. 1991; Reid et al.1995; Kirkpatrick et al. 1999; Martin et al. 1999; Cruz et al. 2002), radial velocities measurement (Bochanski et al. 2007b), metallicity estimation (Gizis 1997; L\'{e}pine et al. 2003; Woolf et al. 2006; L\'{e}pine et al. 2007) and an analysis of magnetic activity (Reid et al. 1995; Hawley et al. 1996; Gizis et al. 2000; West et al. 2004; west et al. 2011).

Because of the difficulty of obtaining spectra from these faint objects, studies on M dwarfs of a decade ago were limited by the number of M dwarf spectra (e.g. Delfosse et al. 1998, 1999, with 118 M stars). However, with the development of modern astronomical facilities, the number of M dwarf spectra increases dramatically. Reid et al. (1995) obtained a spectroscopy catalog of 1746 stars, containing primarily M dwarf spectra and a small number of $ A \sim K $ spectra. SDSS later sharply expanded the number of M dwarfs spectroscopic sample. West et al. (2008) presented a spectroscopic catalog of more than 44,000 M dwarfs from the Sloan Digital Sky Survey (SDSS) Data Release 5 and then as a part of the SEGUE (the Sloan Extension for Galactic Understanding and Exploration) survey (Yanny et al. 2009), over 50,000 additional M dwarf candidates provided a new sight for probing the structure, kinematics, and evolution of the Milky Way. West et al. (2011) presented the latest spectroscopic catalog including 70841 M dwarfs spectra from SDSS Data Release 7 (Abazajian, K. et al. 2009), providing fundamental parameters of M dwarfs for future use of the sample to probe galactic chemical evolution.

The Large Sky Area Multi-Object Fiber Spectroscopic Telescope (LAMOST, also called Guo Shou Jing Telescope) is a National Major Scientific Project undertaken by the Chinese Academy of Science (Wang et al. 1996, Cui et al. 2012). The survey contains two main parts: the LAMOST ExtraGAlactic Survey (LEGAS), and the LAMOST Experiment for Galactic Understanding and Exploration (LEGUE) survey of Milky Way stellar structure. LAMOST has a field of view as large as 20 square degrees, and at the same time a large effective aperture that varies from 3.6 to 4.9 meters in diameter. A limiting magnitude that LAMOST can reach is as faint as r = 19 at resolution R = 1800. This telescope therefore has a great potential to efficiently survey a large volume of space for stars and galaxies. From October 2011 to May 2012, LAMOST has completed the pilot survey and released 319000 spectra (Luo et al. 2012). From the fall of 2012, LAMOST has begun the general survey. Within 4-5 years, LAMOST will observe at least 2.5 million stars in a contiguous area in the Galactic halo, and more than 7.5 million stars in the low galactic latitude areas around the plane. The spectra collected for such a huge sample of stars will provide a legacy that allow us to learn detailed information on stellar kinematics, chemical compositions well beyond SDSS/SEGUE. There is a detailed description of LAMOST spectra survey in Zhao, et al (2012) and the survey science plan of LEGUE in Deng, et al (2012).

Among the spectra that LAMOST pilot survey obtained, M type spectra account for near ten percent of all stellar spectra (LAMOST 1D pipeline recognized as stellar spectra, regardless of S/N). In terms of this proportion, the total number of M dwarfs spectra in the entire survey will be near one million. Such a big sample will enable a number of research topics in exploring the evolution and the structure of the Milky Way. Because countless molecules bands leads to peculiar morphology of M-type stellar spectra, during the process of LAMOST data analysis, M type spectra have to be treated separately to derive fundamental parameters, while other type stellar spectra are input to the ULySS program (Wu et al. 2011) for parameters of temperature, gravity and metal abundance. Our attention is to drive accurate fundamental parameters from M dwarfs spectra obtained in LAMOST pilot survey. This will lay the foundation for further research on the Galactic structure and kinematics.

In this paper we describe the M dwarfs spectra from LAMOST pilot survey and the methods adopted to derive the fundamental parameters including spectral subtype, radial velocity, equivalent width of H${\alpha}$, a number of prominent molecular band indices, the metal sensitive parameter $\zeta$ , and their uncertainties. In section 2 we describe observational spectra from LAMOST and the spectral quality of M dwarfs. In section 3 we discuss how to determine the spectral types for M dwarfs. The radial velocity and its uncertainty of M dwarfs are discussed in section 4. Magnetic activity, molecular band indices and metal sensitive parameter are discussed respectively in section 5. We summarize the results of this paper in section 6.

\section{LAMOST pilot data and Observations}
Now the resolution of LAMOST spectra is R=1500 over a wavelength range of $3700 \AA \sim 9000\AA$. Two arms of each spectrograph cover the entire wavelength range with 200 $\AA$ of overlap. The spectral coverage of blue is  $ 3700 \AA \sim 5900\AA$ while that of red is $5700 \AA \sim 9000\AA$. The raw data have been reduced with LAMOST 2D and 1D pipeline (Luo et al. 2004), including bias subtraction, cosmic-ray removal, spectral trace and extraction, flat-fielding, wavelength calibration, and sky subtraction. The red side has higher throughput than the blue. Therefore, it is easy to obtain high quality spectra of M dwarfs than other types of stars because most of the light of M dwarfs is in the red band. Although the latest version of LAMOST pipeline can accurately process the spectra in steps of wavelength calibration and flat-fielding, there is still some uncertainty in the continuum level because of a lack of a high-precision companion photometric survey telescope to aid in absolute flux calibration. In LAMOST sample, most high quality M dwarf spectra matched well with the corresponding templates, but a small fraction of spectra showed different slope than they should, which may caused by interstellar reddening. We need to consider about these factors which could impact the accuracy of our results. To estimate the effect of continuum uncertainties, we compared 348 objects that both LAMOST and SDSS have their spectra. Generally, the S/N of the spectra of these objects in LAMOST are lower than in SDSS. Most of the 348 samples show flux difference smaller than 20\%. Figure 1 shows an example of the comparison. The blue band is from SDSS and the red one is from LAMOST, both are from observing the same object. The two spectra morphologically agree well and the flux difference of two spectra is not more than 10\%.

To separate M dwarfs from other stars, we selected the spectra that were classified as M or K7 by the LAMOST 1D pipeline. LAMOST 1D pipeline carries out chi-squared fits of the observed spectra to the templates. The templates were constructed by linear combinations of eigen spectra (from decomposition of a set of SDSS spectra) and low-order polynomials (Luo et al. 2012). Most M-type stellar spectra were correctly recognized by the 1D pipeline, but near one fifth M-type stellar spectra (mainly early-type M) were misclassified as K7 dwarfs. Therefore, all K7 type spectra need to be inspected to search for M-type spectra. The total number of the candidates is 98,887.  After visually inspection using the Hammer spectral typing facility (Covey et al. 2007), we excluded K7 type spectra, double stars and poor quality spectra. We ignored giants contamination because giant contamination rate is very low (Covey et al. 2008b). Our final LAMOST catalog contains 67,082 M dwarfs.

It is important to note that not all spectra of this M dwarf sample are in the released stellar spectra from LAMOST pilot survey. The released stellar spectra need meet S/N larger than 10 (in the g and r bands).  But for many M-type stellar spectra, even with a lower S/N (in r band), they can be easily identified and assigned with an appropriate spectral type. Thus, we collected all M-type spectra into this M dwarf sample irrespective of S/N.

\section{SPECTRAL TYPES}
Spectral subtype of an M dwarf is one of the most important fundamental parameters, which relates to the temperature and the mass of the M dwarf. There are two primary approaches to classify a M dwarf according its spectrum. The one uses the overall slope of the spectrum, which requires accurate spectrophotometric calibration over the full optical wavelength. The other matches the relative strength of atomic and molecular features in the spectrum which has been normalized by dividing by an estimated continuum. Considering the flux uncertainty of LAMOST M dwarf spectra, we choose the second method to derive the spectral types of the LAMOST M dwarfs.

The Hammer (Covey et al. 2007) is an IDL-based code that uses the relative strength of features. It has been widely used for classifying stellar spectra (Lee et al. 2008a; West et al. 2011; Woolf et al. 2012; Dhital et al. 2012), especially for M dwarf spectra. For late-type stars classification, the Hammer computes 16 molecular band head indices of each spectrum, including indices of CaI, MgI, CaH3, TiO5, VO, NaI, Cs, CrH, CaII, etc. The wavelength of the indices covers from $4000\AA$ to $9100\AA$.  The Hammer matches these indices with the indices computed from templates, and the spectral type of the closest template is selected as the spectral type of the observed spectra. However, as described in the previous findings, the automatic Hammer tends to classify some later-than-M5 spectra as an earlier subtype (West et al. 2011).

To try to remedy the Hammer of the late M classification problem, we first tested the Hammer code with known type M dwarf spectra. Bochanski et al (2007b, here after Bochanski2007b) derived low-mass M0$\sim$L0 template spectra that were computed from over 4000 SDSS spectra, and used them for medium-resolution radial velocity standards. We adopted Bochanski2007b template to test the Hammer automatic spectral typing. We used the Hammer automatically classify the Bochanski2007b templates and the results are shown in the second column of Table 2. In contrast of the first column of the table, M1, M5, M6, M7 and M9 template are allocated with an earlier subtype.

We inspected each mismatch case and found that the spectral region between 6000\AA\ and 7000\AA\ and the one beyond 8000\AA\ are not well matched between the Bochanski2007b template and the Hammer template. It means that the original Hammer indices are not adequate to discriminate all of the subtypes. We run an ensemble learning method Random Forest (Breiman, 2001) in search of the most important features for classifying M dwarf spectra.

Random forest is an ensemble classifier, which consists of many decision trees and aggregates their results. The method injects randomness to guarantee trees in the forest are different. This somewhat counterintuitive strategy turns out to perform very well comparing to many other classifiers, including discriminant analysis, support vector and neural networks (Liaw et al. 2002). Random forests have become increasingly popular in many scientific fields (C.Strobl et al. 2008). And variable importance measures of random forest have been receiving increased attention as a method of features selection in many classification tasks in bioinformatics and related scientific fields (D\'{i}az-Uriarte R. et al.2006, C.Strobl et al. 2007 ).

Before we classified Bochanski2007b M dwarf template using random forest, we divided each spectrum from 6000\AA{ } to 9000\AA{ }into 600 regions, each region covers 5\AA. The adopted five pseudo continuum regions are: 6130\AA $\sim$ 6134\AA{ }, 6545\AA $\sim$ 6549\AA{ }, 7042\AA$\sim$ 7046\AA{ }, 7560\AA $\sim$ 7564\AA{ }, 8125\AA $\sim$ 8129\AA{ }. Mean flux of each region from 600 regions was computed and then was divided by the mean flux of each pseudo continuum. Finally, 3000 indices were obtained for each spectrum. These indices and corresponding subtypes were input to construct the forest. Knowledge or patterns were learnt from the inputs during construction so that the constructed forest can classify a spectrum according to what it learnt. Simultaneously the variable importance measurements were provided by the forest.

The important features (the numerators of important indices) supplied by Random Forest method are shown in Figure 2. As a large amount of LAMOST spectra do not have high quality in the blue part, and most of obvious features of M dwarfs that can help to distinguish the spectral subtypes locate in 6000$\sim$ 9000\AA{ }, the spectral range we care about is limited to 6000\AA $\sim$ 9000\AA{ }. According to the features importance list, we adjusted the indices of the Hammer by adding three new indices and keeping the original Hammer indices that are in 6000$\sim $9000\AA{ }. The three indices and their corresponding wavelength ranges are shown in Table 1. CaH6385, TiO8250 are sensitive to temperature. The range of index 6545 is near H${\alpha}$ and is often used as a part of continuum to compute the equivalent width of H${\alpha}$. So we take the index 6545 as a pseudo continuum index, and named it Color6545. Table 1 shows the wavelength ranges of each index. The original Hammer indice CaI, MgI and Color1 are removed because they are out of the wavelength of 6000$\sim $9000\AA{ }. After adjusting indices of the Hammer, there are still 16 indices. The 16 indices values of Hammer M dwarf template are calculated again. Figure 3 showed the values of recomputed indices and three extra indices. In Figure 3, the points in the same color are indices values in the same subtype. Lines from red to blue are corresponding to the subtypes of M0 to M9. This figure indicates which indices are good in distinguishing subtypes of M dwarf spectra.

We then classified the Bochanski2007b template using the modified Hammer. The results are showed in the third column of Table 2. The classification result of each template is right now. In order to verify the performance of Hammer after adjusting indices, we further tested the code with 70,841 spectra from the SDSS DR7 M dwarf catalog (West et al 2011). All spectral subtypes in the catalog were derived by visual inspection. The results are showed in Figure 4. The left panel of the figure is the differences distribution of all spectral subtypes for total 70,841 spectra. It shows a number of spectra that were classified a later subtype by the original Hammer, are partially corrected by the modified Hammer. The accuracy of the modified Hammer is higher than the original Hammer. The right panel shows the differences distribution of subtypes for later than M5 spectra and indicates the original Hammer classifies a larger fraction late-type M dwarfs as an earlier subtype (as indicated by West et al. 2011). This is greatly remedied by the improved Hammer. According to the statistical results of all the data, the mean subtype difference is 1 subtype before adjusting indices while the mean subtype difference is 0.6 subtypes for modified Hammer.

We used the amended Hammer to classify the 67,082 M dwarf spectra from LAMOST pilot survey. Figure 5 shows the signal to noise (S/N) distribution of M dwarfs from LAMOST pilot survey. The S/N was computed in the range of 6900\AA $\sim$ 8160\AA. Figure 6 shows the spectral subtypes distribution of M dwarfs in this catalog. From this distribution, early-type M dwarfs account for a large proportion, and the number of late-type M dwarfs from M6 to M9 is small (only 724). This is likely due to target selection effects and the resolution capacity limitation of LAMOST telescope.

\section{Radial Velocity}
The radial velocity (RV) of each M dwarf was measured by the cross-correlation method. Each observed spectrum was cross-correlated with the Bochanski2007b M dwarf template of best matched subtype. In order to decrease the impact of inaccuracy of flux calibration, we used a cubic polynomial to rectify the observed spectra to best fitted template spectra. We constrained the range of radial velocities to $\pm$500km/s, then moved the observed spectra from -500km/s to 500km/s in a 2km/s step. After each move, the observed spectra was multiplied an optimal cubic polynomial to cross-correlate the observed spectrum, with the corresponding template fitting all correlation values to produce a Gaussian peak. The corresponding radial velocity of the Gaussian peak was chosen as our final radial velocity. A bootstrap estimate was conducted to access the internal error of radial velocity estimation.

We used spectra from the SDSS DR7 M dwarf catalog to test our RV measure method. Figure 7 shows the results. The left panel shows the RV comparison to the West2011 values. The right panel shows the distribution of RV differences, in which 67843 RVs differences between -100km/s and +100km/s are shown. Figure 7 indicates that the RVs we measured generally agree with the RVs West et al. measured. The mean of two RVs difference is 0.17km/s while the standard deviation is 6.4km/s, which is less than the reported uncertainties of West2011. The larger scatter of RV around the center of the figure is due to low S/N of spectra, which can be seen in Figure 8. It is intrinsically difficult to derive accurate RVs from these spectra with low S/N. There are a group of points in the bottom of the left panel of Figure 7. We visually inspected the Na doublet and found the RVs of these spectra measured by West2011 have larger uncertainties. A Na doublet at 8183\AA{ }and 8195\AA{ } fitting example is showed in the Figure 9, in which template spectrum was plotted in red and observed spectrum was plotted in blue. The observed spectrum was corrected to zero radial velocity by RV values respectively from west2011 (top panel) and from our method (bottom panel). We further selected a subsample to check the performance of our method. In this subsample, the S/Ns are between 10 and 20 and the differences of two RVs are between 50 and 200. Of total 479 spectra from this sample, for about 180 spectra, our RV is better than West's, and about 104 worse than West's and for the rest 195 spectra, the accuracy of two RVs is the same.

There are many factors which may cause radial velocity uncertainties. The uncertainty of using cross-correlation method is due to some reasons such as the resolution of the spectra, signal noise ratio, the accuracy of the wavelength calibration and flux calibration,  and matching the spectral type of the template to the observed spectra. We used the best matched spectral type template and cross-correlated it with the observed spectra, which minimizes the error introduced by spectral type mismatch. For flux calibration problem, LAMOST has relative flux calibration instead of absolute calibration, which may lead to inaccuracy of spectral flux, and for the wavelength calibration, the RV error caused by inaccurate wavelength calibration of LAMOST spectra is less than 10km/s (Luo2012).

We computed RV and error for all spectra of our catalog and got a mean internal error 11.5 km/s. All spectra of this catalog then were corrected using our radial velocity value for further measurements of H${\alpha}$ emission line and molecular band indices.

\section{MAGNETIC ACTIVITY \& MOLECULAR BAND INDICES}
H${\alpha}$ emission line is the best indicator of chromospheric magnetic activity in M dwarfs due to their red colors. We estimated the magnetic activity of M dwarfs according to the methods of West et al. (2004, 2011). We used a total 14$\AA$ wavelength region for calculation of equivalent width of H${\alpha}$. The central wavelength is 6564.66\AA{ }in vacuum with 7\AA{ }on either side. The continua regions are 6555.0\AA$ \sim$ 6560.0\AA{ } and 6570.0\AA$\sim$6575.0\AA{ }. Our magnetic activity criteria are similar to the west2011 criteria. As our sample contained all spectra of M dwarf from the pivot survey irrespective of S/N, we add an additional S/N criterion (5) to obtain a more clean activity sample. Our criteria is (1) The S/N of continuum near H${\alpha}$ is larger than 3, (2) the EW of H${\alpha}$ must larger than 1, (3) EW is larger than three times of error, (4) the height of the emission line is larger than three times of noise in the adjacent continuum, (5) the S/N of 6500$\sim$6550\AA{ } and 6575\AA$\sim$6625\AA{ } is larger than 10. A star is classified as inactive, it should meet the criteria of (1) and (5) mentioned before to insure the spectrum has higher S/N, besides the spectrum has no detectable emission.

Using the criteria, from 67082 M dwarf spectra, 2312 of them are H${\alpha}$ active while 26074 are H${\alpha}$ inactive, the H${\alpha}$ activity fraction of M0 to M5 listed in Table 3. We confirm that later subtypes have higher active fractions and the trend of the active fraction from M0 to M5 is in agreement with West2011. But in this sample, the number of late-type M dwarfs from M6 to M9 is 724, and only 174 spectra with S/N$>$10. This is likely due to target selection effects and the resolution capacity limitation of LAMOST telescope. The number of late-type M dwarfs is too small to produce a reasonable activity fraction if using the same magnetic activity criteria. Therefore the activity fraction of M6$ \sim $ M9 was not provided here.

We have also computed important molecular band features TiO1$\sim$TiO5, CaH1$\sim$CaH3, and CaOH according to the wavelength ranges defined by Reid et al (1995a). The errors of these indices are given as well.

A rough indicator of metallicity $\zeta$ was computed, which was defined by L\'{e}pine et al. (2007). This $\zeta$ based on the strength of TiO5, CaH2 and CaH3 molecular bands. According to the indicator M spectra can be divided into different metallicity classes: dwarf, subdwarf, extreme dwarf and ultra subdwarfs. However, L\'{e}pine et al. (2012) found $\zeta$ are not sensitive enough to diagnose metallicity variations in dwarfs of subtypes M2 and earlier. Dhital et al. (2012) refined the indice $\zeta$ and the new $\zeta$ can better fit their observed M sample, in which most are M0$\sim$M3 dwarfs spectra. We measured the parameter $\zeta$ and its error according to the refined definition of $\zeta$ by Dhital et al.. Note that Mann et al. (2013) further tested the $\zeta$ parameter with their sample and found $\zeta$ correlates with [Fe/H] for super-solar metallicities, but $\zeta$ does not always correctly identify metal-poor M dwarfs.

\section{SUMMARY}
We present an M dwarf spectral catalog from LAMOST pilot survey, which consists of 67082 M dwarfs. In this catalog, spectral subtypes, radial velocities, magnetic activity, equivalent widths of H${\alpha}$ and various molecular band indices are provided. Spectral subtypes have been derived by the remedied Hammer program and then confirmed by visually inspection. The amended Hammer results have an average offset of 0.6 subtype compared to the visually inspection results. The radial velocities of the sample have been measured with cross-correlation method, with an average 11.5km/s internal error. We also estimate the magnetic activity of M dwarfs through measuring the equivalent width and strength of H${\alpha}$.

The M dwarf catalog of LAMOST pilot survey provides the first glance of LAMOST M-type stellar spectra and it is the first step of our research. Our subsequent work will focus on obtaining more information about M dwarfs, such as kinematics, metallicity, distance and mass through cross-matching with other catalogs. The formal LAMOST survey will enlarge the sample of M dwarfs rapidly and even get the largest spectroscopic sample of M dwarfs. Large samples can enable more and scientific studies of M dwarfs and provide more statistically significant results to explore the structure and evolution of the Milky Way.

\acknowledgments
 We would like to thank Dr. Wei Du for helpful discussions toward the implementation and development of data process software and Ms. Min-Yi, Lin for continuing help with astronomical research methods. This research had made use of LAMOST data and SDSS data. This research is supported by the National Natural Science Foundation of China(Grant Nos 10973021,11078013,10978010,11078019 and 11263001).

\clearpage

\begin{figure}
\includegraphics[angle=0,scale=1]{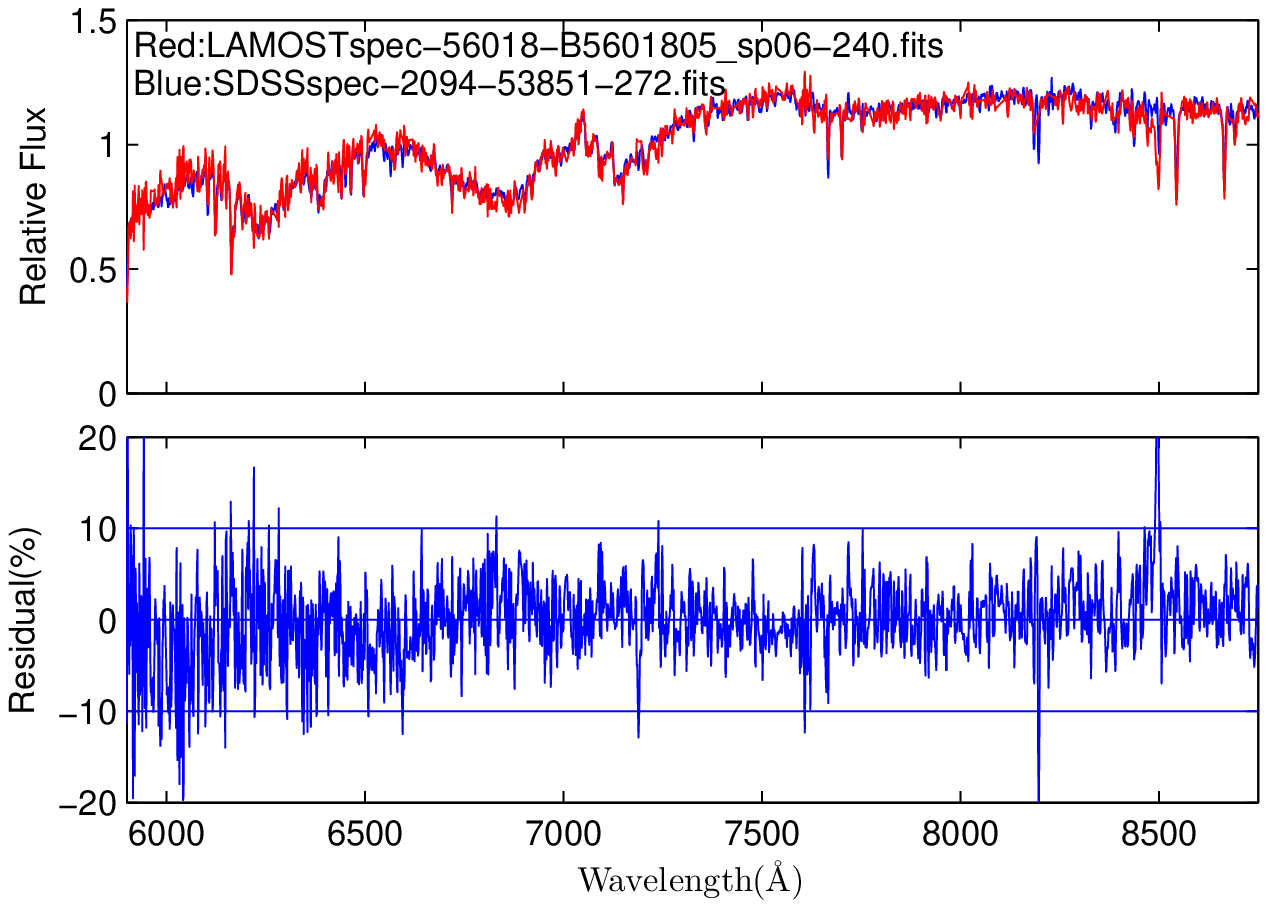}
\caption{Comparison between LAMOST and SDSS spectra, top panel shows two spectra with SDSS spectra in blue and LAMOST spectra in red; bottom panel is the difference of the two spectra.}
\end{figure}

\clearpage

\begin{figure}
\includegraphics[angle=0,scale=1]{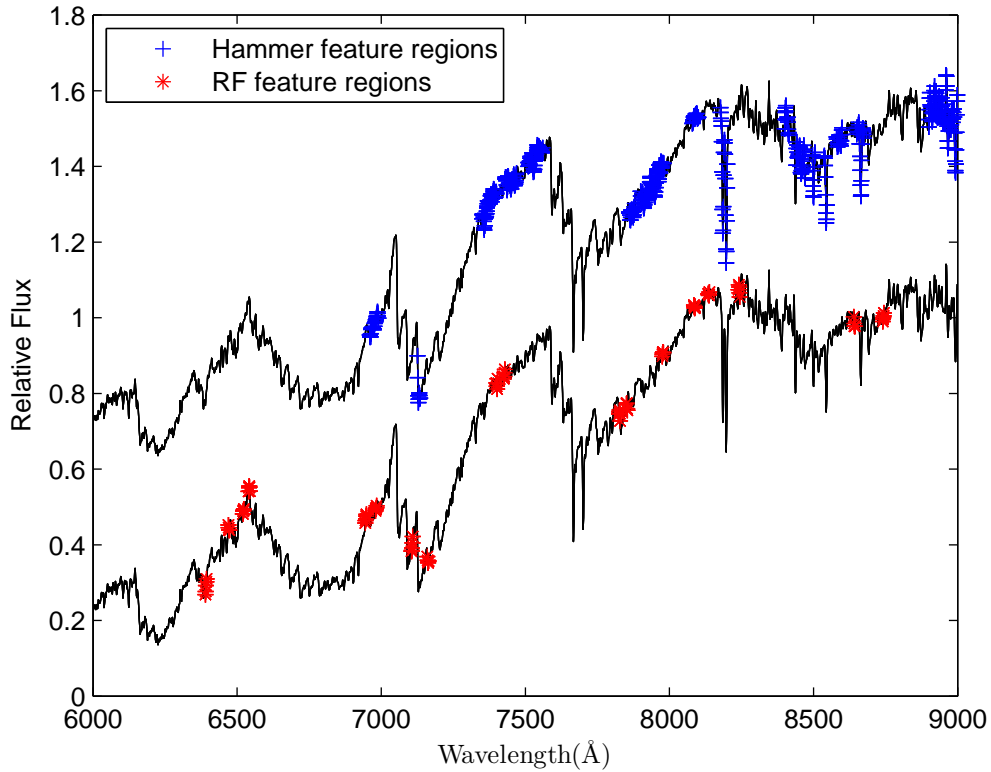}
\caption{The Hammer/Random Forest features. The two spectra are the same M4 dwarf spectrum from Bochanski2007b template. Top: The blue crosses identify regions that are used by the Hammer ,18 total regions.  Bottom: The 18 most important features, which represent the numerators of the important indices that the random forest recommends, are highlighted in red stars. }
\end{figure}
\clearpage

\begin{figure}
\begin{center}
\centering
\includegraphics[angle=270,scale=0.65]{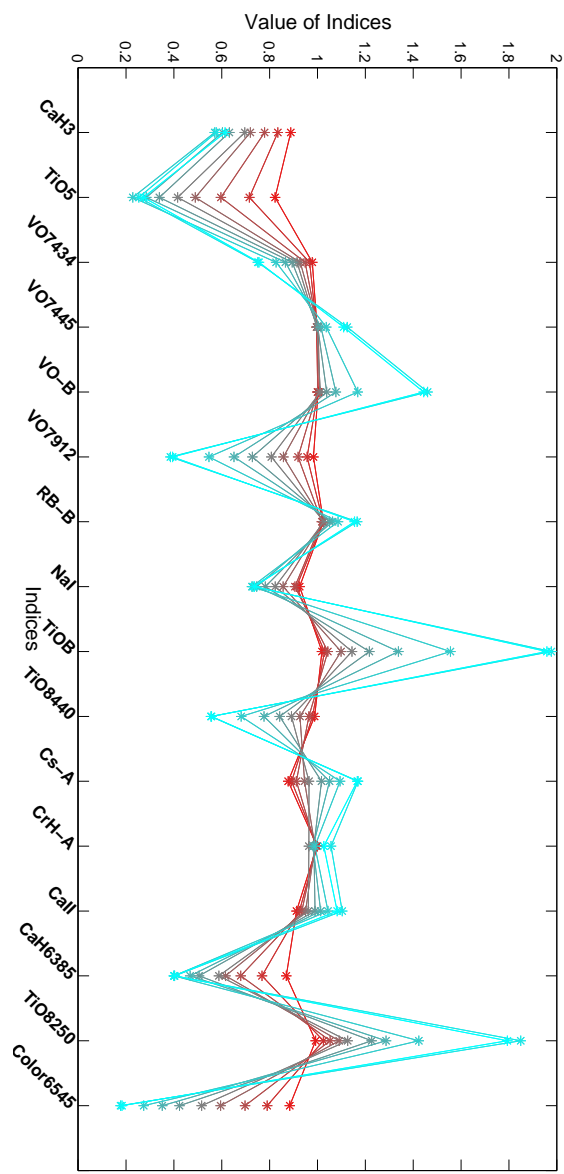}
\caption{The modified Hammer indices, including 13 Original Hammer indices in $6000 \sim 9000$\AA  and newly added three indices. The new-added three indices: CaH6385, TiO8250,Color6545. Each line corresponds to a different subtype. Lines from red  to blue standards for subtypes from M0 to M9. If certain index value increases or reduces monofonically with the type from M0 ~ M9, it indicates that the index is a good spectral type tracer and it is suitable for spectrum type classification. For example, VO7912, TiOB, TiO8250 and Color6545 are good spectral type tracers.}
\end{center}
\end{figure}
\clearpage

\begin{figure}
\includegraphics[angle=0,scale=.70]{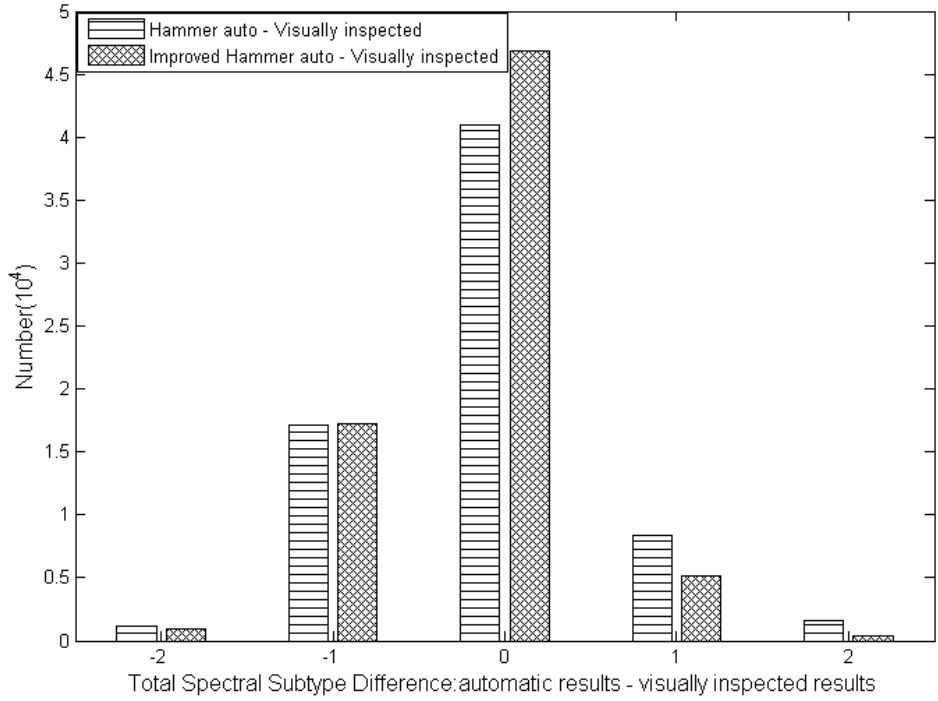}\hspace{10mm} \includegraphics[angle=0,scale=.70]{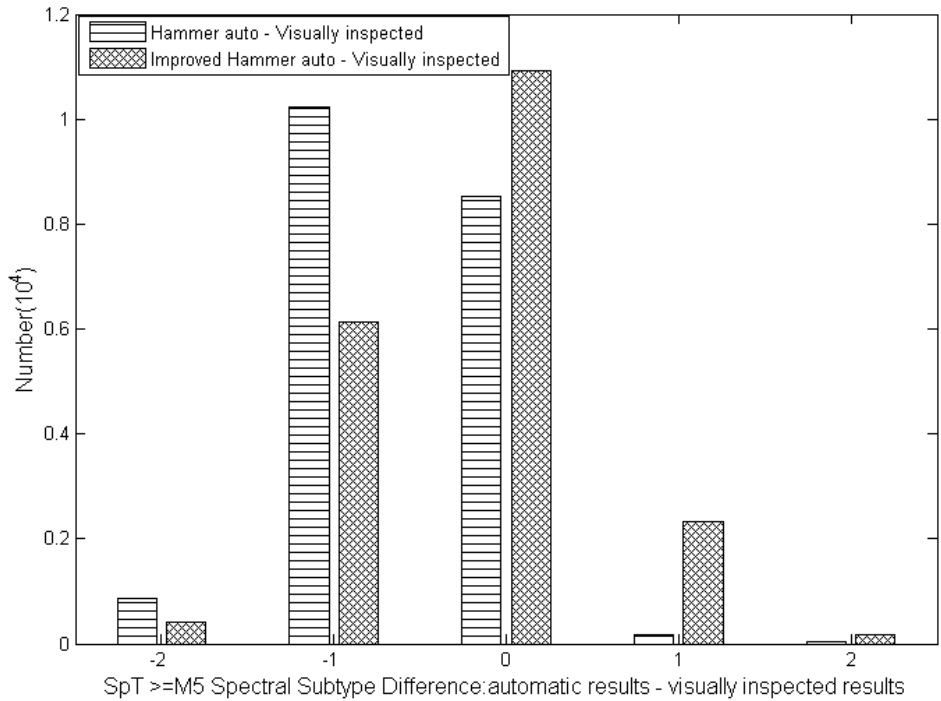}
\caption{Distribution of differences between Hammer/improved Hammer and visually inspected spectral subtype for M dwarfs. The top panel is the distribution of differences for total 70841 spectra(West et al.2011). The bottom panel is the distribution of differences for later than M5 spectra.}
\end{figure}

\clearpage
\begin{figure}
\includegraphics[angle=0,scale=.90]{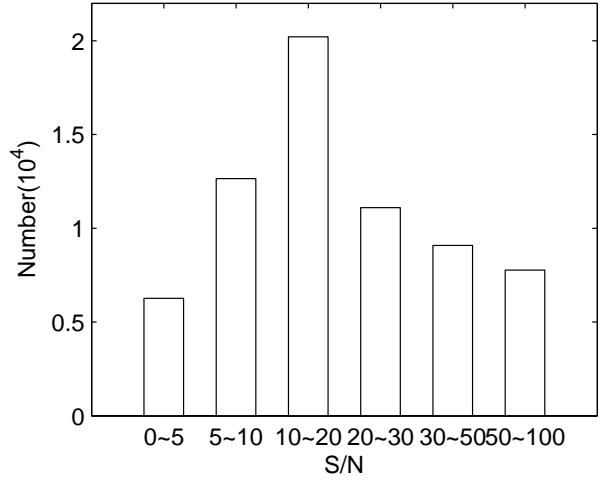}
\caption{S/N distribution of M dwarfs in the LAMOST pilot survey }
\end{figure}

\begin{figure}
\includegraphics[angle=0,scale=.80]{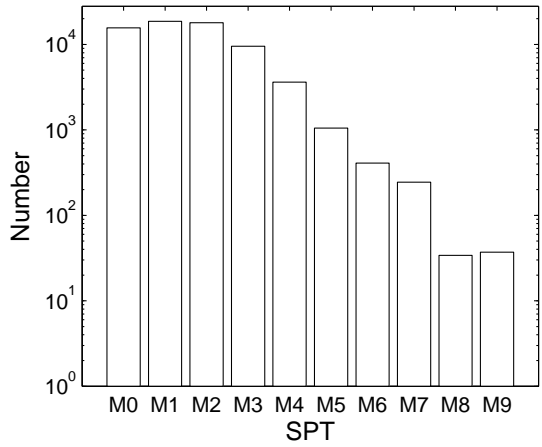}
\caption{Spectral subtype distribution of M dwarfs in the LAMOST pilot survey }
\end{figure}

\clearpage
\begin{figure}
\includegraphics[angle=0,scale=.60]{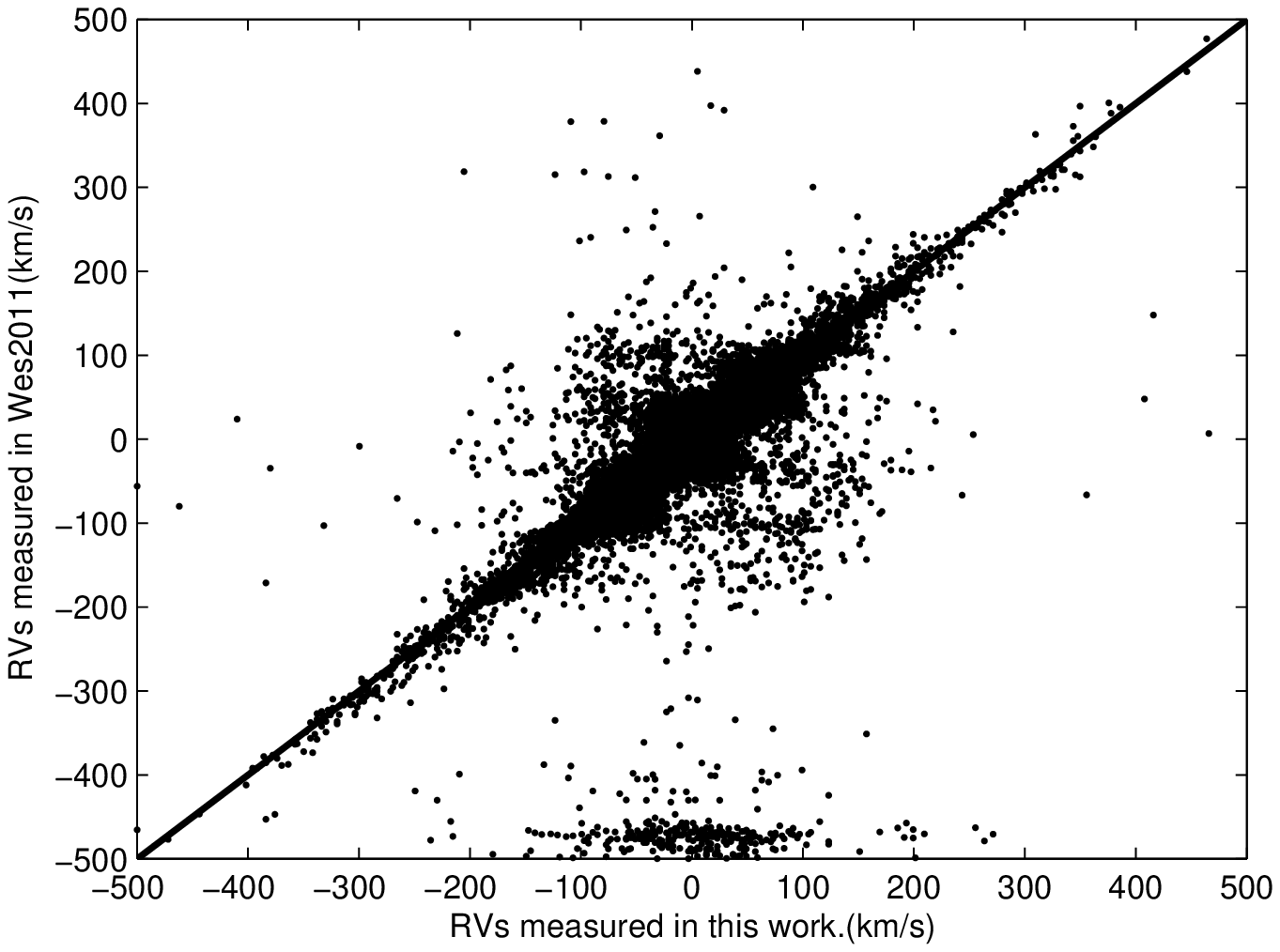}\hspace{1mm} \includegraphics[angle=0,scale=.60]{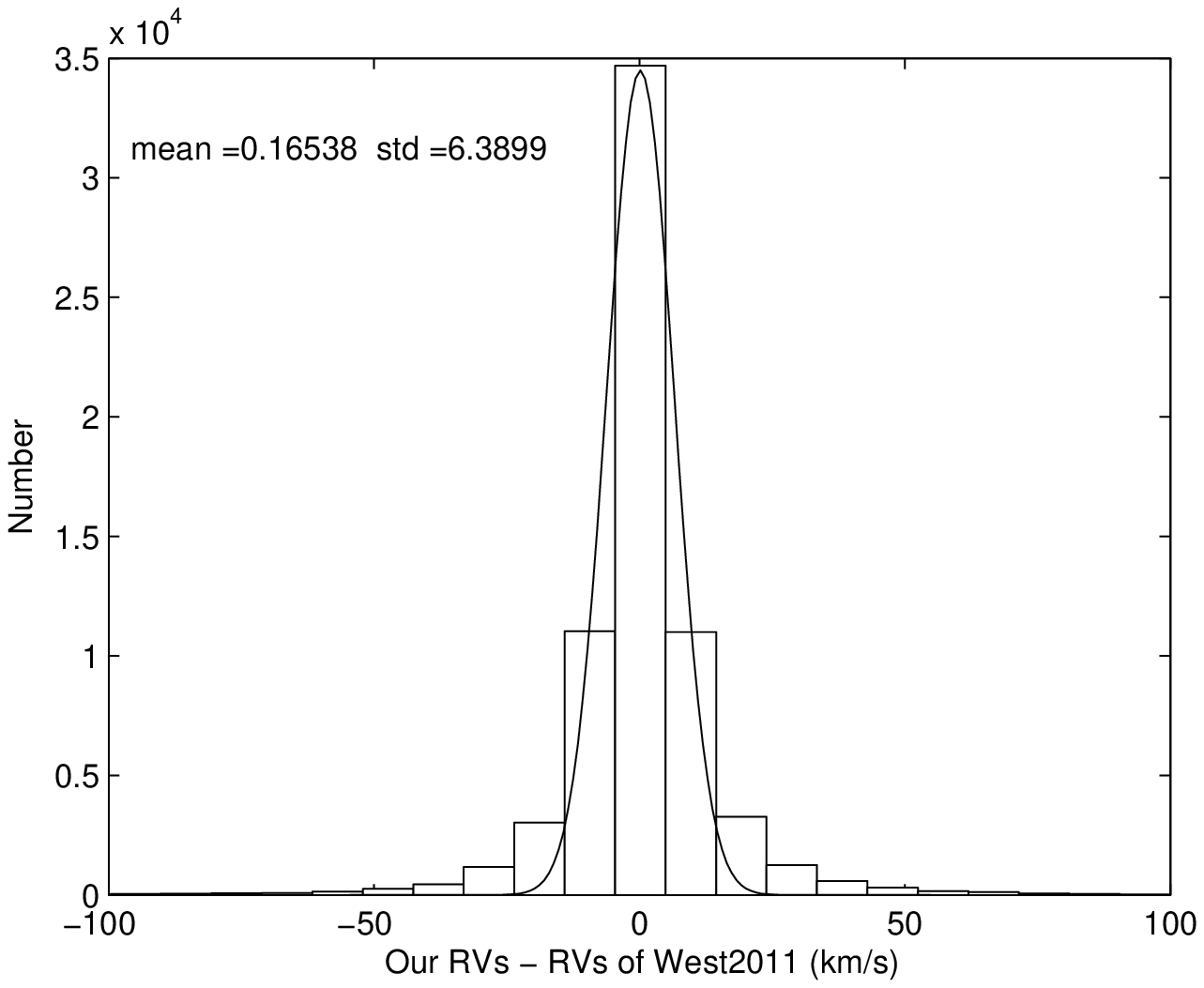}
\caption{RV comparison. The left panel shows the comparison between RVs measured using our method and the RVs measured by West. The right is the distribution of RVs differences.}
\end{figure}

\begin{figure}
\includegraphics[angle=0,scale=.70]{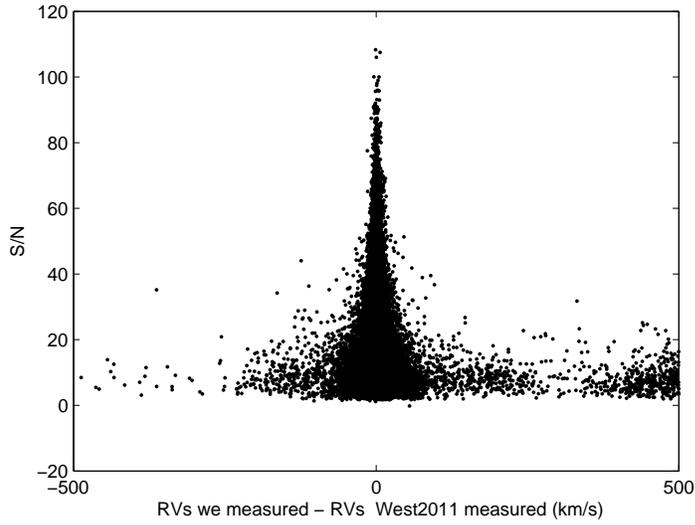}
\caption{The relation of differences of RVs as a function of S/N. When S/N is less than 20, there is more scatter in the differences in the RV values.}
\end{figure}

\clearpage
\begin{figure}
\includegraphics[angle=0,scale=.80]{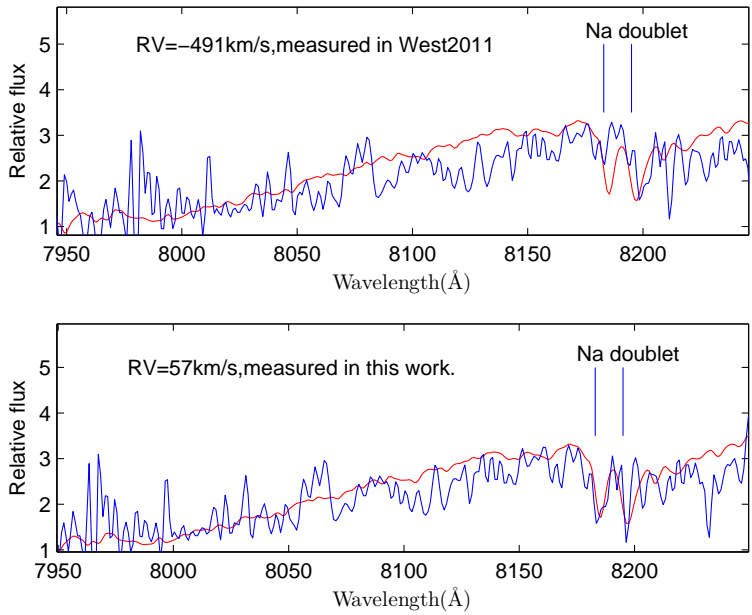}
\caption{Example of Na doublet fitting at  8183 \AA and 8195 \AA. The template spectra are plotted in red and the observed spectra are plotted in blue. The observed spectra was corrected to zero radial velocity respectively according to the RV values measured by west2011 (top panel) and  by our method (bottom panel) }
\end{figure}

\clearpage
\begin{deluxetable}{ccrrrrrrrrcrl}
\tablecaption{New-added Three Indices}
\tablewidth{0pt}
\tablehead{
\colhead{Index} & \multicolumn{2}{c}{$\lambda_w$[\AA]} & \multicolumn{2}{c}{$\lambda_c$$_o$$_n$$_t$[\AA]}
}
\startdata
CaH6385  & 6385.0 & 6389.0 & 6545.0 & 6549.0 \\
TiO8250 & 8250.0 & 8254.0 & 7560.0 & 7564.0 \\
Color6545	 & 6545.0 & 6549.0 & 7560.0 & 7564.0 \\
\enddata
\end{deluxetable}

\begin{table}
\begin{center}
\tabletypesize{\scriptsize}
\caption{Original Hammer and improved Hammer classification of  Bochanski2007 M dwarf templates}
\begin{tabular}{cccc}
\tableline\tableline
Bochanski2007b  & Original Hammer  & Improved Hammer  & Eyecheck \\
Subtype & Autotype & Autotype &  Subtype\\
\tableline
M0 & M0 & M0 & M0\\
M1 & M0 & M1 & M1\\
M2 & M2 & M2 & M2\\
M3 & M3 & M3 & M3\\
M4 & M4 & M4 & M4\\
M5 & M4 & M5 & M5\\
M6 & M5 & M6 & M6\\
M7 & M6 & M7 & M7\\
M8 & M8 & M8 & M8\\
M9 & M8 & M9 & M9\\
\tableline
\end{tabular}
\end{center}
\end{table}

\clearpage
\begin{table}
\begin{center}
\caption{H${\alpha}$ activity fraction on each subtype}
\begin{tabular}{ccccc}
\tableline\tableline
SPT & Total & Active & Inactive & Active Fraction  \\
\tableline

 M0 & 15657 & 267 & 10408 & 2.50\\
    M1 & 18594 & 471  & 7805  & 5.69\\
    M2 & 17941 & 650 & 5231 & 11.05\\
    M3 & 9500 & 600 & 1970 & 23.35\\
    M4 & 3619 & 258 & 479 & 35.01 \\
    M5 & 1047 & 53 & 82 & 39.26 \\
\tableline
\end{tabular}
\end{center}
\end{table}
							
\clearpage



\begin{thebibliography}{}
\bibitem[{{Abazajian} \& {Sloan Digital Sky Survey}(2009)}]{abazajian2009}{Abazajian}, K. \& {Sloan Digital Sky Survey}, f.~t. 2009, \apjs, 182, 543
\bibitem[Bochanski(2011)]{Bochanski2011} Bochanski, J. J., Hawley, S. L., \& West, A. A. 2011, AJ, 141, 98
\bibitem[Bochanski(2010)]{Bochanski2010} Bochanski, J. J., Hawley, S. L., Covey, K. R., et al. 2010, AJ, 139, 2679
\bibitem[Bochanski(2005)]{Bochanski2005} Bochanski, J. J., Hawley, S. L., Reid, I. N., et al. 2005, AJ, 130, 1871
\bibitem[Bochanski(2007a)]{Bochanski2007a} Bochanski, J. J., Munn, J. A., Hawley, S. L., et al. 2007a, AJ, 134, 2418
\bibitem[Bochanski(2007b)]{Bochanski2007b} Bochanski, J. J., West, A. A., Hawley, S. L., \& Covey, K. R. 2007b, AJ, 133,531
\bibitem[Breiman(2001)]{Breiman2001} Breiman, L.,2001, Machine Learning,45,5.
\bibitem[Burgasser(2002)]{Burgasser2002} Burgasser, A. J., Liebert, J., Kirkpatrick, J. D., \& Gizis, J. E. 2002, AJ, 123,2744
\bibitem[Charbonneau(2009)]{Charbonneau2009} Charbonneau, D., et al. 2009, Nature, 462, 891
\bibitem[Covey(2007)]{Covey2007} Covey, K. R., et al. 2007, AJ, 134, 2398
\bibitem[Covey(2008a)]{Covey2008a} Covey, K. R., et al. 2008a, ApJS, 178, 339
\bibitem[Covey(2008b)]{Covey2008b} Covey, K. R., et al. 2008b, AJ, 136, 1778
\bibitem[Cruz(2002)]{Cruz2002} Cruz, K. L., Reid, I. N.,2002,AJ, 123, 2828
\bibitem[Cruz(2003)]{Cruz2003} Cruz,K. L., Reid, I. N., Liebert, J., Kirkpatrick, J.D.,\&Lowrance, P. J. 2003,AJ,126, 2421
\bibitem[Cui(2012)]{Cui2012} Cui, X.Q., Zhao, Y.H., Chu, Y.Q. et al. 2012, Research in Astron. Astrophys. (RAA), 12, 1197
\bibitem[Delfosse(1999)]{Delfosse1999} Delfosse, X., Forveille, T., Beuzit, J.-L., et al. 1999, A\&A, 344, 897
\bibitem[Delfosse(1998)]{Delfosse1998} Delfosse, X., Forveille, T., Perrier, C., \& Mayor, M. 1998, A\&A, 331, 581
\bibitem[Deng(2012)]{Deng2012}Deng, L. C., et al. 2012, Research in Astron. Astrophys. (RAA), 12, 735
\bibitem[Dhital(2012)]{Dhital2012} DHITAL, S., West, A. A., Stassun, K. G., et al. 2012, AJ, 143, 67
\bibitem[D\'{i}az-Uriarte(2006)]{Diaz-Uriarte2006}D \'{i}az-Uriarte R. \& S. Alvarez de Andr\'{e}s, 2006, BMC Bioinformatics 7, 3
\bibitem[FC Adams(2004)]{Adams2004} FC Adams, G Laughlin, \& G J.M Graves, 2004, RevMexAA, 22, 46
\bibitem[Gizis(1997)]{Gizis1997} Gizis, J. E., 1997, AJ, 113, 806
\bibitem[Gizis(2002)]{Gizis2002} Gizis, J. E., Reid, I. N., \& Hawley, S. L. 2002, AJ, 123, 3356
\bibitem[Gizis(2000)]{Gizis2000} Gizis, J. E., Monet, D. G., Reid, I. N., et al. 2000, AJ, 120, 1085
\bibitem[Hawley(1991)]{Hawley1991} Hawley, S. L., \& Pettersen, B. R. 1991, ApJ, 378, 725
\bibitem[Hawley(2002)]{Hawley2002} Hawley, S. L., et al. 2002, AJ, 123, 3409
\bibitem[Hawley(1996)]{Hawley1996} Hawley, S. L., Gizis, J. E., \& Reid, I. N. 1996, AJ, 112, 2799
\bibitem[Kerber(2001)]{Kerber2001} Kerber, L. O., Javiel, S. C., \& Santiago, B. X. 2001, A\&A, 365, 424
\bibitem[Kirkpatrick(1999)]{Kirkpatrick1999} Kirkpatrick, J. D., et al. 1999, ApJ, 519, 802
\bibitem[Kirkpatrick(1991)]{Kirkpatrick1991} Kirkpatrick, J. D., Henry, T. J., \& McCarthy, D. W. 1991, ApJS, 77, 417
\bibitem[Laughlin(1997)]{Laughlin1997} Laughlin, G., Bodenheimer, P., \& Adams, F. C. 1997, ApJ, 482, 420
\bibitem[Lee(2008a)]{Lee2008a} Lee, Y. S., et al. 2008a, AJ, 136, 2022
\bibitem[L\'{e}pine(2008)]{Lepine2008} L\'{e}pine, S., \& Scholz, R. 2008, ApJ, 681, L33
\bibitem[L\'{e}pine(2005)]{Lepine2005} L\'{e}pine, S., \& Shara, M. M. 2005, AJ, 129, 1483
\bibitem[L\'{e}pine(2007)]{Lepine2007} L\'{e}pine, S., Rich, R. M., \& Shara, M. M. 2007, ApJ, 669, 1235
\bibitem[L\'{e}pine(2003)]{Lepine2003} L\'{e}pine, S., Rich, R. M., \& Shara, M. M. 2003, AJ, 125, 1598
\bibitem[L\'{e}pine(2012)]{Lepine2012} L\'{e}pine, S., Hilton, E. J., Mann, A. W., et al. 2012, arXiv:1206.5991
\bibitem[Liaw(2002)]{Liaw2002} Liaw, A., \& Wiener, M., 2002, R News, 2:18-22
\bibitem[Luo(2004)]{Luo2004} Luo, A. L., Zhang, Y. X., \& Zhao, Y. H. 2004, Proc. SPIE, 5496, 756
\bibitem[Luo(2012)]{Luo2012} Luo, A. L., Zhang, H. T., \& Zhao, Y. H. 2012, Research in Astron. Astrophys. (RAA), 12, 1243
\bibitem[Mann(2013)]{Mann2013}Mann, A. W., Brewer, J. M., Gaidos, E., et al.,2013,AJ,145,52
\bibitem[Martin(1999)]{Martin1999} Martin, E. L., 1999, MNRAS, 302, 59
\bibitem[Reid(1997)]{Reid1997} Reid, I. N., Gizis, J. E., Cohen, J. G., et al. 1997, PASP, 109, 559
\bibitem[Reid(1995a)]{Reid1995a} Reid, I. N., Hawley, S. L., \& Gizis, J. E. 1995a, AJ, 110, 1838
\bibitem[Reid(1995b)]{Reid1995b} Reid, N., Hawley, S. L., \& Mateo, M. 1995b, MNRAS, 272, 828
\bibitem[Schmidt(2010a)]{Schmidt2010a} Schmidt, S. J., West, A. A., Burgasser, A. J., et al. 2010a, AJ, 139, 1045
\bibitem[Schmidt(2010b)]{Schmidt2010b} Schmidt, S. J., West, A. A., Hawley, S. L., \& Pineda, J. S. 2010b, AJ, 139,1808
\bibitem[Strobl(2007)]{Strobl2007} Strobl,C., Boulesteix, A.-L., Zeileis, A. \& Hothorn T. 2007, BMC Bioinformatics, 8, 25
\bibitem[Strobl(2008)]{Strobl2008} Strobl,C., Boulesteix, A.-L., Kneib, T., Augustin,T. \& Zeileis A. 2008, BMC Bioinformatics, 9, 307
\bibitem[Walkowicz(2009)]{Walkowicz2009} Walkowicz, L. M., \& Hawley, S. L. 2009, AJ, 137, 3297
\bibitem[Wang(1996)]{Wang1996} Wang, S. G., Su, D. Q., Chu, Y. Q., et al. 1996, Appl. Opt., 35, 5155
\bibitem[West(2009)]{West2009} West, A. A., \& Basri, G. 2009, ApJ, 693, 1283
\bibitem[West(2008)]{West2008} West, A. A., \& Hawley, S. L. 2008, PASP, 120, 1161
\bibitem[West(2006)]{West2006} West, A. A., Bochanski, J. J., Hawley, S. L., et al. 2006, AJ, 132, 2507
\bibitem[West(2004)]{West2004} West, A. A., et al. 2004, AJ, 128, 426
\bibitem[West(2011)]{West2011} West, A. A., et al. 2011, AJ, 141, 97
\bibitem[West(2008)]{West2008} West, A. A., Hawley, S. L., Bochanski, J. J., et al. 2008, AJ, 135, 785
\bibitem[Woolf(2006)]{Woolf2006} Woolf, V. M., Wallerstein G., 2006, PASP, 118, 218
\bibitem[Woolf(2012)]{Woolf2012} Woolf, V. M., West, A. A.,2012,MNRAS, 422, 1489
\bibitem[Yanny(2009)]{Yanny2009} Yanny, B., et al., 2009, AJ, 137, 4377
\bibitem[Zhao(2012)]{Zhao2012} Zhao G., et al., 2012, Research in Astron. Astrophys. (RAA), 12,723
\end{thebibliography}
\end{document}